\documentclass[%
 aip,
 jmp,%
 amsmath,amssymb,
 reprint,%
]{revtex4-1}

\usepackage{graphicx}
\usepackage{dcolumn}
\usepackage{bm}

\begin{document}

\preprint{AIP}

\title{Giant Linear Magneto-resistance in Nonmagnetic PtBi$_2$}

\author{Xiaojun Yang}
 \affiliation{Department of Physics and State Key Laboratory of Silicon Materials, Zhejiang University, Hangzhou 310027, China}
\author{Hua Bai}
 \affiliation{Department of Physics and State Key Laboratory of Silicon Materials, Zhejiang University, Hangzhou 310027, China}
\author{Zhen Wang}
 \affiliation{Department of Physics and State Key Laboratory of Silicon Materials, Zhejiang University, Hangzhou 310027, China}
\author{Yupeng Li}
 \affiliation{Department of Physics and State Key Laboratory of Silicon Materials, Zhejiang University, Hangzhou 310027, China}
\author{Qian Chen}
 \affiliation{Department of Physics and State Key Laboratory of Silicon Materials, Zhejiang University, Hangzhou 310027, China}
\author{Jian Chen}
 \affiliation{Department of Physics and State Key Laboratory of Silicon Materials, Zhejiang University, Hangzhou 310027, China}
\author{Yuke Li}
 \affiliation{Department of Physics, Hangzhou Normal University - Hangzhou 310036, China}
\author{Chunmu Feng}
 \affiliation{Department of Physics and State Key Laboratory of Silicon Materials, Zhejiang University, Hangzhou 310027, China}
\author{Yi Zheng}
 \affiliation{Department of Physics and State Key Laboratory of Silicon Materials, Zhejiang University, Hangzhou 310027, China}
 \affiliation{Zhejiang California International NanoSystems Institute, Zhejiang University, Hangzhou 310058, China}
 \affiliation{Collaborative Innovation Centre of Advanced Microstructures, Nanjing 210093, China}
\author{Zhu-an Xu}
 \email{zhuan@zju.edu.cn}
 \affiliation{Department of Physics and State Key Laboratory of Silicon Materials, Zhejiang University, Hangzhou 310027, China}
 \affiliation{Zhejiang California International NanoSystems Institute, Zhejiang University, Hangzhou 310058, China}
 \affiliation{Collaborative Innovation Centre of Advanced Microstructures, Nanjing 210093, China}
\date{\today}

\begin{abstract}
We synthesized nonmagnetic PtBi$_2$ single crystals and observed a
giant linear magneto-resistance (MR) up to 684\% under a magnetic
field $\mu_0H$ = 15 T at $T$ = 2 K. The linear MR decreases with
increasing temperature, but it is still as large as 61\% under
$\mu_0H$ of 15 T at room temperature. Such a giant linear MR is
unlikely to be described by the quantum model as the quantum
condition is not satisfied. Instead, we found that the slope of MR
scales with the Hall mobility, and it can be well explained by a
classical disorder model.

\end{abstract}

\pacs{75.47.Gk; 72.15.Gd; 71.20.Lp; 85.75.Bb}

\maketitle

Materials exhibiting large magneto-resistance (MR) can not only be
utilized to enlarge the sensitivity of read/write heads of
magnetic storage devices, e.g., magnetic memory\cite{LMR_memory}
and hard drives\cite{LMR_hard driver}, but also stimulate many
fundamental studies in material physics at low
temperatures\cite{LMR_funderment,WTe2_nature}. Generally speaking,
the ordinary MR in non-magnetic compounds and elements\cite{weak
effect_science} is a relatively weak effect and usually at the
level of a few percent for metals\cite{WTe2_nature}. Moreover, a
conventional conductor under an applied magnetic field exhibits a
quadratic field dependence of MR which saturates at medium fields
and shows a relatively small magnitude. Owing to the rich physics
and potential applications, the large linear MR effect has drawn
renewed interest recently.\cite{rni_nature,rni_NC,APL_classical}
There are two predominant models used to explain the origin of
such large linear MR effect, namely, the quantum
model\cite{theory_AAA} and the classical model\cite{cla did
nature}. The quantum model is proposed for materials with zero
band gap and linear energy dispersion, such as topological
insulators\cite{QTI_PRL}, graphene\cite{QGraphe_NL}, Dirac
semimetals like SrMnBi$_2$\cite{QSrMnBi2_PRL}, and the parent
compounds of iron based superconductors \cite{QBaFe2As2_PRL}.
Quantum linear MR occurs in the quantum limit when all of the
electrons fill the lowest Landau level (LL)\cite{theory_AAA}. In
contrast, the classical linear MR is dominated by disorder.
Materials showing the classical linear MR include highly
disordered systems\cite{cla dis PRB}, and weakly disordered
samples with high mobility\cite{InSb_NM,Cd3As2_PRL,cla hig
JAP,APL_classical}, thin films, and quantum Hall systems
\cite{CquanHall_PRL}. However, it is interesting that the
classical linear MR has also frequently been reported in materials
with linear dispersions, such as the topological insulator
Bi$_2$Se$_3$ \cite{Bi2Se3_APL}, graphene\cite{APL_classical}, and
the Dirac semimetal Cd$_3$As$_2$\cite{Cd3As2_PRL}, which may be
due to their large mobility. Even weak disorder could induce
linear MR in high-mobility samples\cite{cla hig JAP,Cd3As2_PRL}.
When the carrier concentration is too high for the quantum limit,
the linear MR may be described by classical model for disordered
systems\cite{InSb_NM,APL_classical}.

In this Letter, we synthesized high quality single crystals of
nonmagnetic PtBi$_2$ and investigated the magnetotransport
properties. We observed a giant positive linear MR up to 684\%
under $\mu_0H$ = 15 T at $T$ = 2 K. MR decreases with increasing
temperature, but MR of 61\% is still achieved under a magnetic
field of 15 T even at room temperature. Regarding the origin of
the linear MR, the close relationship between the MR and the Hall
mobility implies that the observed linear MR should not be
attributed to the quantum origin, but may be explained by the
classical model.

The PtBi$_2$ single crystals were synthesized using a self-flux
method. Powders of the elements Pt (99.97\%) and Bi (99.99\%),
both from Alfa Aesar, were thoroughly mixed together in an atomic
ratio of Pt:Bi = 1:8, before being loaded into a small alumina
crucible. The crucible was then sealed in a quartz tube in Argon
gas atmosphere. During the growth, the quartz tube was slowly
heated up to 1273 K and kept at the temperature for 10 h. Finally
it was slowly cooled to 873 K at a rate of -3 K/h, followed by
centrifugation to remove the excessive Bi. The resulting single
crystals are large plates with a typical dimension of 3 $\times$ 3
$\times$ 0.05 mm$^3$. We cut the single crystal into a rectangle
of about 1 $\times$ 0.4 $\times$ 0.05 mm$^3$ for transport
measurements. The stoichiometry and structure of these single
crystals were checked using Energy-dispersive X-ray spectroscopy
(EDX) and X-ray diffraction (XRD) measurements. All transport
measurements were carried out in an Oxford-15 T cryostat with a
He4 probe in a Hall-bar geometry, using Keithley 2400 sourcemeters
and 2182A nanovoltmeters.

\begin{figure}
\includegraphics[width=8cm]{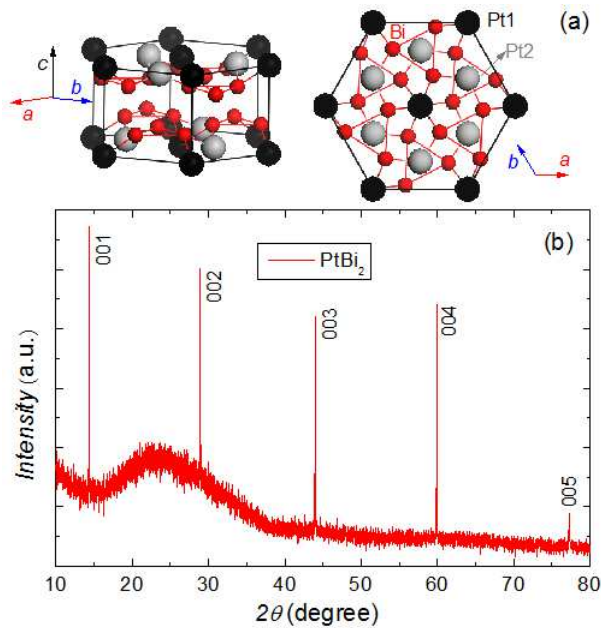}
\caption{\label{Fig.1} (color online)  (a), The crystal structure of PtBi$_2$.
(b), X-ray diffraction patterns of PtBi$_2$ single-crystal sample.}
\end{figure}

As illustrated in Fig. 1(a), PtBi$_2$ has a layered pyrite crystal
structure with the space group of P-3 (No.
147)\cite{structure_PtBi2}. Fig. 1(b) shows the X-ray diffraction
pattern of the PtBi$_2$ single crystals. Only multiple reflections
of (0 0 $l$) planes can be detected, consistent with the layered
crystal structure depicted in Fig. 1(a). The interplane spacing is
determined to be 6.16 {\AA}, agreeing well with the previous
reported value\cite{structure_PtBi2}.

\begin{figure}
\includegraphics[width = 8cm]{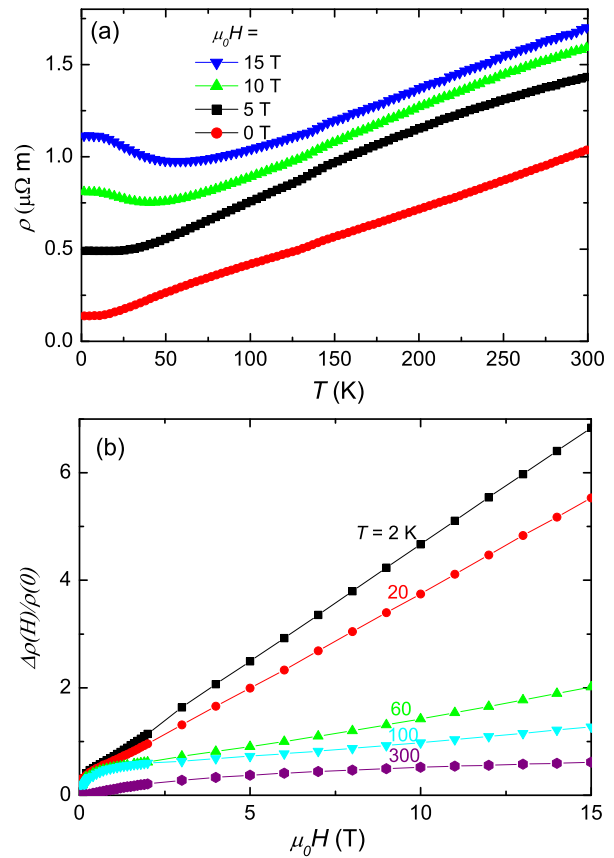}
\caption{\label{Fig.2}(color online)  (a), In-plane resistivity
$\rho$ as a function of temperature $T$ at a series of
out-of-plane magnetic fields $\mu_0H$ = 0, 5, 10 and 15 T. (b),
The magnetoresistance (MR = $\Delta \rho(H)/\rho(0) =
\rho(H)/\rho(0) - 1$) as a function of magnetic field $\mu_0H$ at
a series of temperatures $T$ = 2, 20, 60, 100 and 300 K.}
\end{figure}

The temperature dependence of in-plane resistivity curves under
magnetic fields ($||c$) of $\mu_0H$ = 0, 5, 10 and 15 T are
displayed in Fig. 2(a). In zero field, the room temperature
resistivity is 1.1 $\mu\Omega$ m and decreases to 0.13 $\mu\Omega$
m at  2 K, yielding a residual resistivity ratio (RRR) of 8.5.
When a field is applied, the resistivity increases rapidly,
corresponding to a large positive magnetoresistance (MR = $\Delta
\rho(H)/\rho(0) = \rho(H)/\rho(0) - 1$). By measuring the magnetic
field dependence of resistivity at fixed temperatures, we have
observed giant, non-saturating linear MR, which can reach 684\%
under $\mu_0H$ = 15 T at $T$ = 2 K [Fig. 2(b)]. To our surprise,
the room temperature magnetoresistance is still as large as 61\%
in a field of 15 T.

The non-saturating, large linear magnetoresistance in PtBi$_2$ is
quite unusual and contradicts with the semiclassical transport
theory. For conventional metals, the MR exhibits quadratic
field-dependence in the low field range and saturates under high
field, and the MR is usually of a small value. For a system with
open orbits or Fermi surfaces, unsaturated MR with quadratic field
dependence (or linear field dependence, which critically depends
on the Fermi surface and the relative orientation of magnetic
field) could appear even under high field along the open orbits
while in other directions MR would still show saturated behavior.
As a result, linear field dependence of MR could be observed in
polycrystal sample owing to averaging
effect\cite{linearMR_open,LaAgSb2,cyclotron theory}. Such a giant,
non-saturating linear MR in the PtBi$_2$ single crystal certainly
does not fit into these two categories.

\begin{figure}
\includegraphics[width=8cm]{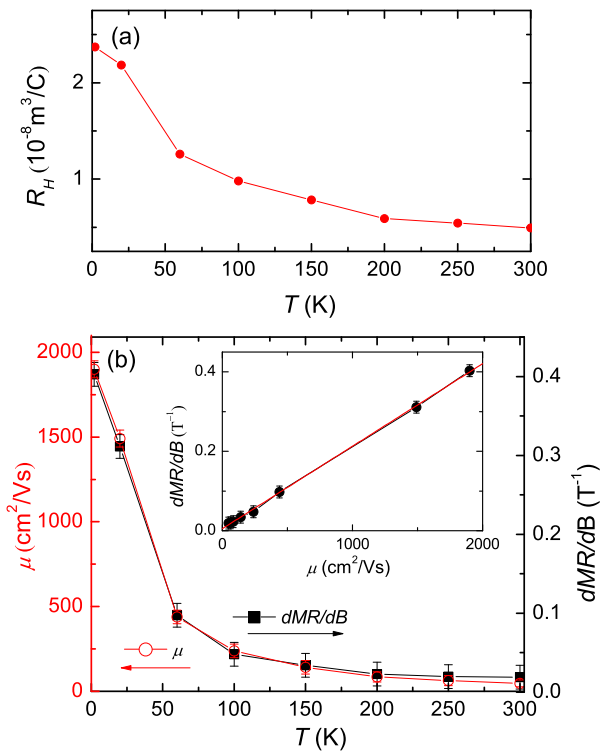}
\caption{\label{Fig.3}(color online)  (a), Hall coefficient versus temperature for PtBi$_2$.
(b), Hall mobility (red hollow circles) and dMR/d$B$ (black solid squares) versus temperature.
The inset displays the dMR/d$B$ versus Hall mobility. }.
\end{figure}

We first consider a quantum explanation for the observed linear MR
phenomenon in the framework developed by
Abrikosov\cite{theory_AAA}. Following this theory, linear MR will
appear in the quantum limit, when $\hbar\omega_c$ exceeds the
Fermi energy $E_F$ and all the electrons occupy the lowest LL. In
such a limit, the quantum magnetoresistivity is calculated as
$\rho_{xx} = N_i B/ \pi n^2 e$, where $n$ and $N_i$ are the
electron density and the concentration of scattering centers,
respectively. The equation is valid under the condition, $n \ll
(eB/\hbar)^{3/2}$. That is, the quantum linear MR would appear
when $B \gg (\hbar /e)n^{3/2}$. In Fig. 3(a), we have plotted the
temperature dependence of Hall coefficient ($R_H$). The Hall
resistivity ($\rho_{yx}$) is linearly dependent on the magnetic
field (not showing here) and thus we use a single-band model to
estimate the charge carrier density, i.e, $n = 1/eR_H$ and to
calculate the critical magnetic field of $(\hbar /e)n^{3/2}$. We
find that, even at 2K, it needs $\sim$271 T to satisfy the
quantum condition, which is far higher than the maximum field of
15 T in our experiments. Therefore, the observed linear MR in
PtBi$_2$ is unlikely to be explained by the quantum model.

Instead, the classical disorder models \cite{cla did nature} may
provide a reasonable explanation for the presence of linear MR in
PtBi$_2$. In the disorder network model, the linear MR appears
when the local current density gains spatial fluctuations in both
magnitude and direction, as a result of inhomogeneous carrier or
mobility distribution\cite{cla did nature,Cd3As2_PRL}. Such
classical linear MR phenomena have been observed in various
disordered systems, such as Bi$_2$Se$_3$\cite{Bi2Se3_APL}, n-doped
Cd$_3$As$_2$\cite{Cd3As2_PRL}, and epitaxial graphene on
SiC\cite{APL_classical}. In the classical model, the slope
dMR/d$B$ is predicted to be proportional to the Hall mobility:
dMR/d$B$ $\propto \mu$. Indeed, the dMR/d$B$, which is defined as
the slop of the linearly fitting line of the linear region of MR
at higher fields, exhibits the same temperature dependence (see
the black solid squares in Fig. 3(b)) as the Hall mobility (red
hollow circles). The curve of dMR/d$B$ versus $\mu$ can be fitted
linearly very well, as shown in the inset of Fig. 3(b). This
strongly suggests that the origin of the observed linear MR in our
sample could be classical.

\begin{figure}
\includegraphics[width=8cm]{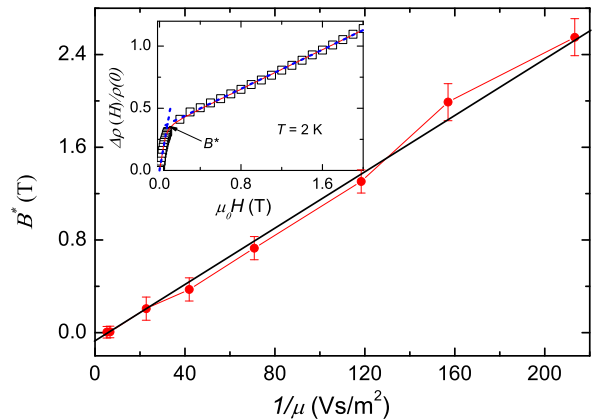}
\caption{\label{Fig.4}(color online)  $B^*$ versus inverse Hall mobility.
Inset: Field dependence of MR (black hollow squares) at $T$ = 2 K.
The two dashed blue lines are the linear fits at the low and high field regimes. The point $B^*$,
marked by the arrow, is defined as the crossing point of these two lines. }.
\end{figure}

This classical origin can be further verified by scaling the inverse Hall
mobility with the crossover magnetic field ($B^*$)\cite{APL_classical}. 
We define the $B^*$, marked by the arrow, as the crossing point of
linear fits at the low and high field regimes, which are shown as
blue dashed lines in the inset of Fig. 4. In the classical model,
the crossover field is predicted to be linearly proportional to
the inverse Hall mobility: $B^* \propto 1/\mu$. Fig. 4 shows $B^*$
versus inverse Hall mobility, which displays good linear
dependence, consistent with the classical model well. This further
confirms that the origin of the observed linear MR could be
classical. The disorder in the single-crystalline samples could
come from the Bi-site vacancies. The occupation of Bi sites
obtained by EDX is around 94\%, which confirms the existence of
Bi-site vacancies.

In summary, we performed detailed magnetotransport property
measurements in the PtBi$_2$ single crystals. PtBi$_2$ exhibits
very large linear magnetoresistance (684\% in 15 T field at 2 K).
The giant linear MR can scale well with the mobility. Our work
indicates that the giant linear MR could arise from the classical
origin, which makes PtBi$_2$ an appealing system for both
practical use and further investigation on its physic properties.

This work is supported by the National Basic Research Program of
China (Grant Nos. 2014CB921203 and 2012CB927404), NSF of China
(Contract Nos. U1332209 and 11190023), the Ministry of Education
of China (Contract No. 2015KF07), and the Fundamental Research
Funds for the Central Universities of China.

\end{document}